\documentclass[reprint,superscriptaddress]{revtex4-2}

\usepackage{graphicx}
\usepackage{bm}
\usepackage{amsmath}
\usepackage{amssymb}
\usepackage{braket}
\usepackage[svgnames]{xcolor}
\usepackage{slashed}
\usepackage[colorlinks=true,allcolors=blue]{hyperref}
\usepackage{cleveref}
\usepackage{subcaption}

\allowdisplaybreaks

\begin{document}

\title{
\mbox{}\Large Next-to-Next-to-Leading Order QCD Corrections to $\eta_t \to HZ$}

\author{Cai-Ping Jia}
\email{jiacp@nwnu.edu.cn}
\affiliation{College of Physics and Electronic Engineering, Northwest Normal University, Lanzhou 730070, China}
\affiliation{Center for High Energy Physics, Peking University, Beijing 100871, China}

\author{Yan-Qing Ma}
\email{yqma@pku.edu.cn}
\affiliation{School of Physics and State Key Laboratory of Nuclear Physics and Technology, Peking University, Beijing 100871, China}
\affiliation{Center for High Energy Physics, Peking University, Beijing 100871, China}

\author{Huai-Min Yu}
\email{huai.min.yu@mib.infn.it}
\affiliation{INFN, Sezione di Milano-Bicocca, Piazza della Scienza 3, 20126 Milano, Italy}

\author{Yu-Jie Zhang}
\email{zyj@buaa.edu.cn}
\affiliation{School of Physics, Beihang University, Beijing 100191, China}

\begin{abstract}
Motivated by recent LHC observations of a threshold enhancement in the $t\bar t$ system consistent with toponium-like dynamics, we compute the next-to-next-to-leading order (NNLO) QCD correction to the hard short-distance coefficient for $\eta_t\to HZ$, where $\eta_t$ denotes the pseudoscalar color-singlet configuration $t\bar t({}^1S_0^{[1]})$.
Within the NRQCD factorization framework, we retain the full dependence on the Higgs- and $Z$-boson masses and include both the two-loop virtual corrections and the real double-gluon-emission channel.
At next-to-leading order (NLO), the finite-mass result remains close to its large-mass limit.
The NNLO coefficient develops logarithms of both the renormalization scale $\mu_R$ and the NRQCD factorization scale $\mu_\Lambda$.
For $\mu_R=\mu_\Lambda=m_t$, the NLO and NNLO terms suppress the leading-order width by about $29\%$ and $19\%$, respectively; the two-loop contribution is about two thirds of the one-loop effect and of the same sign, so the perturbative series converges slowly and the NNLO prediction amounts to about $52\%$ of the leading-order width.
The resulting coefficient provides a necessary ingredient for future threshold studies of $\eta_t\to HZ$ and for assessing the sensitivity of this channel to the top-Higgs interaction.
The accuracy of the result is further supported by three checks: reproduction of the known large-mass limit at one loop, independence of the extracted coefficient on the $\gamma_5$ prescription, and consistency of its scale logarithms with the renormalization-group structure.
\end{abstract}

\maketitle

\section{Introduction}

The top quark plays a special role in the Standard Model. Its mass lies at the electroweak scale, and its Yukawa coupling to the Higgs boson is of order unity. It is therefore a sensitive probe of electroweak symmetry breaking and a natural window on possible new physics~\cite{Djouadi:1996pj,Djouadi:2005gj,Bernreuther:2008ju,Hoang:2020iah,Beneke:2000hk}. At the same time, the top quark has a large decay width and typically decays before hadronization~\cite{Jezabek:1988iv,Jezabek:1993wk,Czarnecki:1997vz,Chetyrkin:1999ju}. This makes top-quark physics perturbatively clean, but also prevents the formation of ordinary narrow quarkonium states.

Nevertheless, the near-threshold $t\bar t$ system remains sensitive to nonrelativistic QCD dynamics. In the region where the relative velocity of the top and antitop is small, Coulomb exchange and threshold resummation can significantly modify both the production rate and the kinematic distributions~\cite{Fadin:1987wz,Fadin:1988fn,Strassler:1990nw,Sumino:1992ai,Hoang:2000yr,Beneke:2015kwa}. Recent LHC measurements have renewed interest in this region. The CMS Collaboration reported a significant excess near the kinematic $t\bar t$ threshold in the dilepton channel. Interpreted with a simplified color-singlet pseudoscalar toponium model, ${}^1S_0^{[1]}$ in spectroscopic notation, the excess cross section was measured to be $8.8^{+1.2}_{-1.4}\ {\rm pb}$, with a statistical significance exceeding five standard deviations~\cite{CMS:2025kzt}. This excess has recently been independently confirmed by CMS in the single-lepton channel~\cite{CMS:2026ani}. The ATLAS Collaboration subsequently observed a compatible effect: models neglecting quasi-bound-state formation are disfavored with a significance of $7.7\sigma$, and the corresponding excess cross section was measured to be $9.0\pm1.3\ {\rm pb}$~\cite{ATLAS:2026nrx}.
These observations provide strong evidence for a threshold-localized enhancement in the $t\bar t$ system, which is consistent with a toponium-like interpretation within simplified models. However, the precise interpretation remains tied to the modeling of the $t\bar t$ threshold region~\cite{Djouadi:2024lyv,Nason:2026oka}. Recent studies have examined the basic properties, effective modeling, collider prospects, and formation dynamics of toponium-like systems~\cite{Fu:2025yft,Fu:2025zxb,Bai:2025buy,Xiong:2025iwg,Tang:2026zhq}.

In this paper we use $\eta_t$ as a shorthand for the pseudoscalar color-singlet configuration $t\bar t({}^1S_0^{[1]})$. This notation does not imply that we treat $\eta_t$ as an ordinary narrow quarkonium resonance~\cite{Fadin:1987wz,Strassler:1990nw,Sumino:1992ai}. Rather, it denotes an on-shell pseudoscalar $t\bar t$ configuration near threshold, whose mass is taken as an external kinematic input.

If the effect is associated with Standard-Model toponium-like dynamics, an important question is whether the same configuration can also lead to signals in independent annihilation channels. The $HZ$ final state is a well-suited candidate for this purpose: it is an electroweak annihilation channel complementary to open-top production, and its rate is directly sensitive to the top-Higgs interaction~\cite{Altenkamp:2012sx,Hespel:2015zea,Hasselhuhn:2016rqt,Chen:2020gae,Wang:2021rxu,Chen:2022rua,Degrassi:2022mro,CampilloAveleira:2025rbh,Davies:2025out,Davies:2026uxl}.
This sensitivity originates from the longitudinal component of the $Z$ boson. Near threshold, the invariant mass of the configuration satisfies $M_{\eta_t}\sim 2m_t \gg m_Z$, and the Goldstone-boson equivalence theorem relates the longitudinal-$Z$ amplitude to the pseudoscalar transition $\eta_t\to H G^0$, where $G^0$ denotes the neutral Goldstone boson~\cite{Lee:1977eg}. Since both $H$ and $G^0$ couple to the top quark through Yukawa interactions, the annihilation amplitude is governed by the top-Higgs coupling. The channel $\eta_t\to HZ$ therefore probes the pseudoscalar heavy-quark configuration and the top-Higgs interaction simultaneously.

If toponium-like dynamics is responsible for the observed threshold excess, the same pseudoscalar configuration should also contribute to $gg\to HZ$ through the chain $gg\to t\bar t({}^1S_0^{[1]})\to HZ$, and a Standard-Model prediction for this contribution requires the corresponding annihilation kernel as precise short-distance input. We therefore compute the QCD corrections to the short-distance annihilation amplitude $t\bar t({}^1S_0^{[1]})\to HZ$, retaining the full dependence on the relevant masses. In a Green-function description, the toponium-like threshold contribution in the pseudoscalar color-singlet channel can be schematically organized as~\cite{Hoang:2000yr,Kiyo:2008bv,Beneke:2011mq,Beneke:2013jia,Beneke:2015kwa,Kawabata:2016aya,Chen:2019fla}
\begin{equation}
\begin{aligned}
{\cal M}_{\rm thr}(gg\to HZ)
\sim\; &
C_{\rm prod}^{({}^1S_0^{[1]})}\,
G(E+i\Gamma_t)\,
C_{\rm ann}^{({}^1S_0^{[1]})}\; .
\end{aligned}
\end{equation}
Here ${\cal M}_{\rm thr}$ denotes only the nonrelativistic threshold contribution mediated by the intermediate state $t\bar t({}^1S_0^{[1]})$, namely ${\cal M}_{\rm thr}(gg\to t\bar t({}^1S_0^{[1]})\to HZ)$, and it should not be understood as the complete fixed-order amplitude for $gg\to HZ$.
The quantity $E$ is the nonrelativistic energy measured relative to the $t\bar t$ threshold, and $\Gamma_t$ denotes the weak decay width of the top quark. The argument $E+i\Gamma_t$ implements the leading finite-width effect of the unstable top quarks and turns the would-be bound-state poles into broadened threshold enhancements. The Green function $G(E+i\Gamma_t)$ describes Coulomb exchange and the resulting threshold line shape, while $C_{\rm prod}^{({}^1S_0^{[1]})}$ and $C_{\rm ann}^{({}^1S_0^{[1]})}$ denote the short-distance coefficients for $gg\to t\bar t({}^1S_0^{[1]})$ and $t\bar t({}^1S_0^{[1]})\to HZ$, respectively. The latter coefficient $C_{\rm ann}^{({}^1S_0^{[1]})}$ is the short-distance object computed in this work.
A full phenomenological description of $gg\to HZ$ would require matching the threshold contribution to the continuum top-loop amplitude and subtracting their common threshold expansion to avoid double counting. If this channel is further connected to the open-top threshold enhancement observed at the LHC, real $t\bar{t}$ production and top-decay effects must also be included consistently. The present work focuses only on the short-distance annihilation part of this problem.

The process $\eta_t\to HZ$ has been considered previously in studies of superheavy pseudoscalar quarkonium decays, where it was identified as an important electroweak annihilation channel. Its QCD correction is known at next-to-leading order (NLO), and related electroweak and Higgs-associated quarkonium decay channels have also been discussed in earlier studies~\cite{Kuhn:1995ee,Djouadi:1997yw,Huang:1996cs,Djouadi:2005gj}. In this work we compute the next-to-next-to-leading order (NNLO) QCD correction to the on-shell annihilation kernel $t\bar t({}^1S_0^{[1]})\to HZ$, retaining the physical $H$- and $Z$-boson masses and the near-threshold pseudoscalar invariant mass. The result provides the hard input needed for future threshold studies of $gg\to HZ$ through a toponium-like intermediate state, and establishes the Standard-Model baseline for the pseudoscalar threshold contribution to the $HZ$ final state.

The rest of this paper is organized as follows. Section~\ref{sec:framework} summarizes the calculation of amplitudes and the renormalization procedure used to extract the finite hard coefficient. Section~\ref{sec:result} presents the NNLO short-distance coefficient and discusses its numerical impact at the level of the hard kernel. Section~\ref{sec:summary} contains our conclusions.

\section{Framework and Methods}\label{sec:framework}

\subsection{Amplitude Calculation}\label{subsec:amp}

Within the NRQCD formalism, the annihilation width for a heavy-quarkonium state can be factorized into a short-distance coefficient and a long-distance matrix element, schematically written as~\cite{Bodwin:1994jh,Petrelli:1997ge}
\begin{equation}
\Gamma(\eta_t\to HZ)=F_{HZ}({}^1S_0)\,\langle \eta_t|\mathcal O({}^1S_0)|\eta_t\rangle+\cdots\;,
\label{eq:fact}
\end{equation}
where $F_{HZ}({}^1S_0)$ is the hard coefficient, and the ellipsis denotes higher-order terms in the velocity expansion. The matrix element encodes the long-distance overlap of the $\eta_t$ state with the nonrelativistic color-singlet $t\bar t$ configuration~\cite{Bodwin:1994jh,Brambilla:2004wf}. Eq.~\eqref{eq:fact} separates hard QCD effects from lower-energy threshold dynamics. In the present work we compute only $F_{HZ}({}^1S_0)$, namely the hard matching coefficient entering this factorized description.

In practical applications it is convenient to absorb the long-distance contribution into the overall normalization and to discuss the perturbative QCD correction factor,
\begin{equation}
K_{HZ}=\frac{\Gamma(\eta_t\to HZ)}{\Gamma_{\mathrm{LO}}(\eta_t\to HZ)}\simeq\frac{F_{HZ}({}^1S_0)}{F_{HZ}^\text{LO}({}^1S_0)}\;.
\label{eq:kdef}
\end{equation}
This separation was already emphasized in the one-loop treatment~\cite{Kuhn:1995ee} and remains essential at two loops~\cite{Hoang:2000yr,Pineda:1997bj,Beneke:2005hg}. 
The hard coefficient is defined as the finite remainder after matching onto the factorized NRQCD description in Eq.~\eqref{eq:fact}.

We compute the NNLO corrections to $\eta_t\to H Z$ starting from the on-shell partonic process
\begin{align}
    t\;(p_1)+\bar{t}\;(p_2)\to H\;(k_1)+Z\;(k_2)\;,
\end{align}
and projecting the initial $t\bar t$ pair onto the color-singlet pseudoscalar configuration ${}^1S_0^{[1]}$. At leading order in the relative-velocity expansion, the corresponding projector is
\begin{equation}
\begin{aligned}\label{eq:projector}
    \mathcal{P}(^1S_0^{[1]})=&\frac{1}{\sqrt{N_c}}\frac{1}{\sqrt{2E}(E+m_t)}\\
    &\times(\slashed{\bar{p}}-m_t)\frac{2E-\slashed{P}}{4E}\gamma_5\frac{2E+\slashed{P}}{4E}(\slashed{p}+m_t)\;,
\end{aligned}
\end{equation}
with kinematics $p=P/2+q$, $\bar{p}=P/2-q$, and $P^2=4m_t^2$. Here $P$ denotes the total momentum of the near-threshold on-shell $t\bar t$ pair, $p$ and $\bar p$ are the momenta of the top and antitop, $q$ is the relative momentum, $E$ is the energy of each heavy quark, and $N_c$ is the number of colors. The relative momentum $q$ is set to zero after the amplitude is expanded to leading order in the nonrelativistic limit.

The perturbative expansion is organized at the level of the projected on-shell partonic width, denoted by $\Gamma$, which is expanded as
\begin{equation}
\begin{aligned}
\Gamma
&=\Gamma_{\rm LO}+\Gamma_{\rm NLO}+\Gamma_{\rm NNLO}+\cdots\\
&\propto\left|\mathcal M^{(0l)}+\mathcal M^{(1l)}+\mathcal M^{(2l)}+\mathcal M^\text{(real)}+\mathcal O(\alpha_s^3)\right|^2\\
&=\mathcal M^{(0l)}\mathcal M^{(0l)*}+
2\,{\rm Re}\!\left[\mathcal M^{(1l)}\mathcal M^{(0l)*}\right]\\
&\quad+\left(2\,{\rm Re}\!\left[\mathcal M^{(2l)}\mathcal M^{(0l)*}\right]+\mathcal M^{(1l)}\mathcal M^{(1l)*}\right)\\
&\quad +\mathcal M^\text{(real)} \mathcal M^\text{(real)*}+\mathcal O(\alpha_s^3)\;.
\end{aligned}
\end{equation}
Here $\mathcal M^{(nl)}$ denotes the $n$-loop virtual amplitude for the two-body channel $\eta_t\to HZ$, while $\mathcal M^{(\mathrm{real})}$ denotes the real-emission amplitudes.
Since the initial $t\bar t$ pair is projected onto a color-singlet state, single-gluon annihilation is forbidden by color conservation, and real radiation therefore starts only at NNLO, through the $HZ+gg$ channel. The $HZ+q\bar q$ channel is absent for the same reason: a light-quark pair produced by a single virtual gluon is equivalent to single-gluon annihilation and is forbidden by the same argument. In a covariant gauge, ghost contributions are included to cancel the unphysical polarizations in the two-gluon channel. Together with the two-loop virtual corrections, this real-emission channel completes the NNLO QCD correction.

Since the spin-singlet pseudoscalar state is projected by a Dirac structure containing $\gamma_5$ in Eq.~\eqref{eq:projector}, while the $Z$ boson couples to the quark current through both vector and axial-vector components, the intermediate Dirac algebra requires a definite prescription for $\gamma_5$ in dimensional regularization. In this work, we carry out the calculation in both the naive anticommuting $\gamma_5$ scheme and the K{\"o}rner--Kreimer--Schilcher (KKS) scheme \cite{Korner:1991sx,Kreimer:1993bh}, thereby providing a nontrivial check on the treatment of the pseudoscalar and axial-vector structures.
In the naive anticommuting scheme, $\gamma_5$ is taken to anticommute with the Dirac matrices in $D=4-2\epsilon$ dimensions, and the usual cyclicity of Dirac traces is retained. In the KKS prescription, $\gamma_5$ likewise anticommutes with the $D$-dimensional Dirac matrices, but for fermion traces containing an odd number of $\gamma_5$ matrices a reading-point prescription must be specified before the trace is reduced to the corresponding four-dimensional Levi-Civita structure~\cite{Korner:1991sx}. In practice, we adopt the symmetrized prescription
\begin{equation}
\mathrm{Tr}(a\cdot \Gamma_J \cdot b)
\;\to\;
\mathrm{Tr}\!\left(\frac{\Gamma_J\cdot b\cdot a + b\cdot a\cdot \Gamma_J}{2}\right)\;,
\end{equation}
where $\Gamma_J$ denotes the insertion containing the unpaired $\gamma_5$, and $a$ and $b$ represent the remaining parts of the fermion chain on the two sides of this insertion~\cite{Tao:2025kqf}.

For the present process, the $\gamma_5$ issue arises only in intermediate Dirac-algebra manipulations of the open quark line and in the axial-current coupling of the $Z$ boson. No anomalous contribution requiring an additional finite axial-current renormalization is encountered in the present calculation. Accordingly, any scheme dependence related to the treatment of $\gamma_5$ is expected to cancel in the renormalized short-distance coefficient. Indeed, after ultraviolet renormalization and IR subtraction described in Sec.~\ref{subsec:renormalization}, we find that the extracted short-distance coefficient is identical in the two schemes. This agreement provides a strong internal consistency check of the calculation.

\begin{figure}[tbp]
  \centering
  \begin{minipage}{0.24\textwidth}
    \centering
    \includegraphics[width=0.8\textwidth]{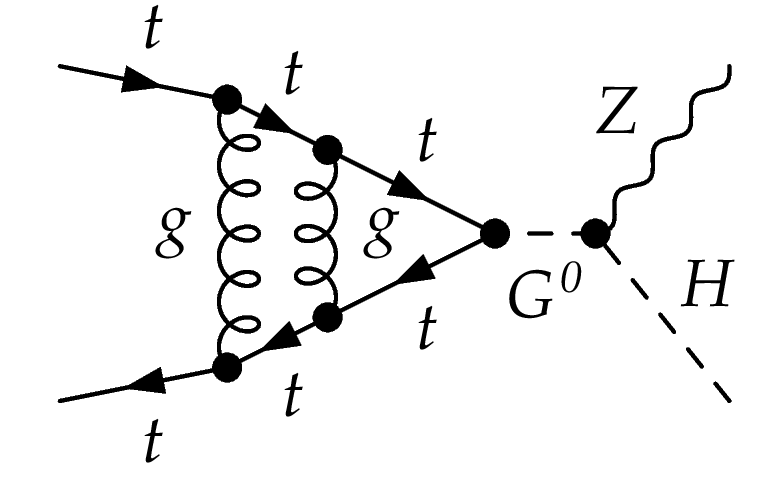}
  \end{minipage}\hfill
  \begin{minipage}{0.24\textwidth}
    \centering
    \includegraphics[width=0.8\textwidth]{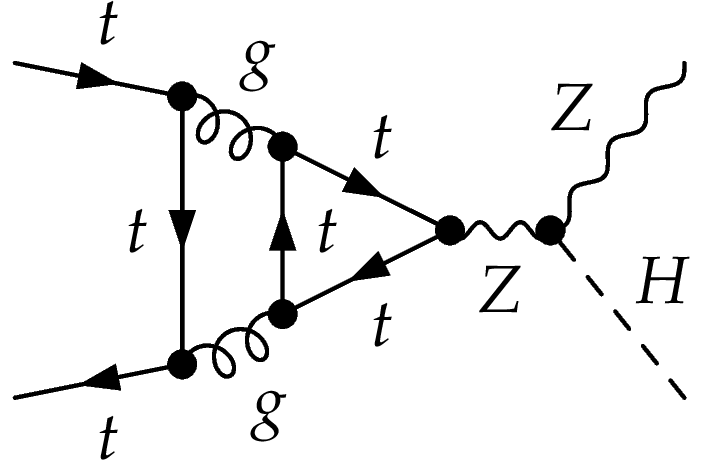}
  \end{minipage}\hfill
  \vspace{2em}
  \begin{minipage}{0.24\textwidth}
    \centering
    \includegraphics[width=0.8\textwidth]{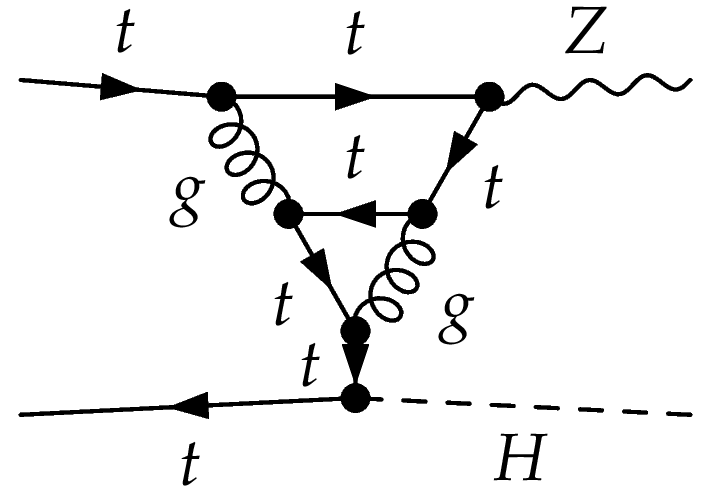}
  \end{minipage}\hfill
  \begin{minipage}{0.24\textwidth}
    \centering
    \includegraphics[width=0.8\textwidth]{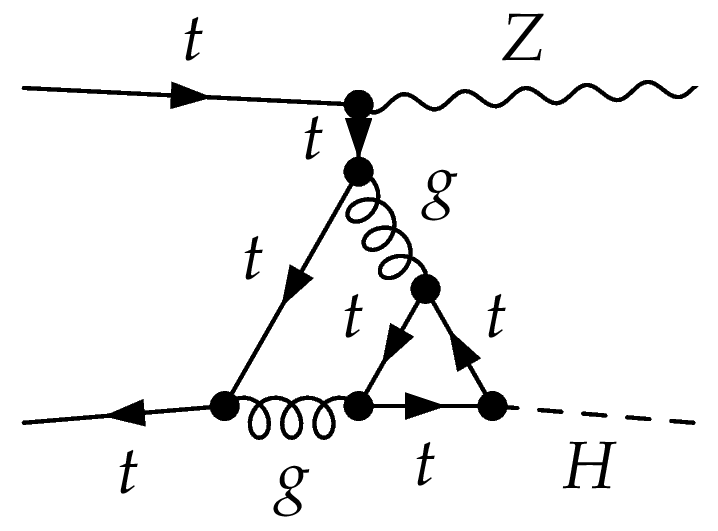}
  \end{minipage}
  \caption{Representative Feynman diagrams for the two-loop virtual QCD corrections to $\eta_t\to HZ$.}
  \label{fig:feynman-topologies}
\end{figure}

Representative Feynman diagrams contributing to $\eta_t\to HZ$ at NNLO are displayed in Fig.~\ref{fig:feynman-topologies}.
All virtual diagrams arise from dressing the heavy-quark lines of the Born-level electroweak diagrams for $t\bar t\to HZ$ with gluons. This organization matches the NRQCD matching picture, in which QCD corrections modify the short-distance annihilation kernel.

The calculation is carried out within an automated framework. We first generate the Feynman diagrams and amplitudes with \texttt{FeynArts}~\cite{Hahn:2000kx}. The spin projection onto the ${}^1S_0^{[1]}$ state, the color-singlet projection, and the subsequent Dirac
and color algebra are performed with \texttt{CalcLoop}~\cite{Ma:calcloop}. After these algebraic manipulations, the squared amplitudes are expressed as linear combinations of scalar integrals belonging to a set of integral families. These integrals are reduced to master integrals with \texttt{Blade}~\cite{Guan:2024byi}, which uses
\texttt{FiniteFlow}~\cite{Peraro:2019svx} to solve the
integration-by-parts identities~\cite{Chetyrkin:1981qh} and exploits the block-triangular form of the IBP system to improve the reduction efficiency~\cite{Liu:2018dmc,Guan:2019bcx}. The master integrals are then evaluated numerically with \texttt{AMFlow}~\cite{Liu:2017jxz,Liu:2021wks, Liu:2022mfb,Liu:2022chg}, which implements the auxiliary-mass-flow method.

\subsection{Renormalization and IR Subtraction}\label{subsec:renormalization}

Following Ref.~\cite{Chen:2025qgy}, the NNLO short-distance coefficient is defined as the finite hard contribution remaining after ultraviolet renormalization and NRQCD matching. We renormalize the heavy-quark mass and field in the on-shell scheme, while the strong coupling is renormalized in the $\overline{\rm MS}$ scheme~\cite{Chetyrkin:1997dh}. The bare and renormalized parameters are related by
\begin{equation}
\begin{aligned}\label{eq:renormalization}
    m_{Q,0} &= Z_m^{\rm OS}\, m_Q\;,\\
    \psi_{Q,0} &= \sqrt{Z_2^{\rm OS}}\,\psi_Q\;,\\
    \alpha_{s,0} &= (4\pi e^{-\gamma_E})^{-\epsilon}\,\mu_R^{2\epsilon}\,
    Z_{\alpha_s}^{\overline{\rm MS}}\,\alpha_s(\mu_R)\;,
\end{aligned}
\end{equation}
where the subscript $0$ denotes bare quantities, $Q$ labels the top quark, $\gamma_E$ is the Euler constant, and $\mu_R$ is the renormalization scale; dimensional regularization is used with $D=4-2\epsilon$. The renormalization constants are taken from Ref.~\cite{Barnreuther:2013qvf}. The electroweak vertices in the Born process $t\bar t\to HZ$ are kept at tree level, since only QCD corrections to the short-distance annihilation kernel are considered. The strong coupling and its renormalization-group evolution are consistently treated in full QCD with six active flavors, $n_f=6$.

After ultraviolet renormalization, the projected partonic amplitude is free of UV divergences, but the virtual correction still contains a residual single pole in $1/\epsilon$. This leftover pole is infrared in origin, reflects the residual long-distance sensitivity of the projected color-singlet amplitude, and is proportional to the same spin-color structure as the leading-order ${}^1S_0^{[1]}$ amplitude. Within the NRQCD factorization framework it is not part of the hard coefficient itself; instead, it is absorbed into the renormalization of the leading four-fermion operator $\mathcal O(^1S_0^{[1]})$.
The corresponding operator renormalization in the $\overline{\rm MS}$ scheme reads~\cite{Chung:2020zqc},
\begin{equation}
\begin{aligned}
&\langle\mathcal{O}^{(Q\bar{Q})[^1S_0^{[1]}]}(^1S_0^{[1]})\rangle|_{\overline{\text{MS}}}\\
=&2N_c\left[1-\alpha_s^2(\mu_R)
\left(\frac{\mu_R^2}{\mu_\Lambda^2}\frac{e^{\gamma_E}}{4\pi}\right)^{2\epsilon}
\left(C_F^2+\frac{C_FC_A}{2}\right)\frac{1}{2\epsilon}\right]\;,
\end{aligned}
\end{equation}
where the operator notation follows Ref.~\cite{Chung:2020zqc}, $C_F=(N_c^2-1)/(2N_c)$ and $C_A=N_c$ are the quadratic Casimir eigenvalues of the fundamental and adjoint color representations, and the scale $\mu_\Lambda$ parameterizes the separation between the hard annihilation kernel and lower-energy nonrelativistic dynamics. Matching the ultraviolet-renormalized QCD amplitude onto the renormalized NRQCD operator removes the remaining infrared pole, and the finite remainder defines the NNLO short-distance coefficient. The logarithmic dependence on $\mu_\Lambda$ appearing in the final hard coefficient is thereby fixed by the factorization procedure.

The real-emission contribution plays a special role at this order: the phase-space-integrated real part is finite and generates no $1/\epsilon$ pole, so the infrared singular structure relevant for extracting the NNLO hard coefficient is entirely contained in the virtual amplitude and is removed by the operator renormalization described above.

\section{Results and Discussion}
\label{sec:result}

Throughout this work, we use the Standard Model input parameters
\begin{equation}\label{eq:input}
\frac{m_H^2}{m_t^2}=\frac{12}{23},\qquad
\frac{m_Z^2}{m_t^2}=\frac{23}{83},\qquad
\frac{m_W^2}{m_t^2}=\frac{14}{65},
\end{equation}
with the top-quark mass chosen as $m_t=172.69\,\mathrm{GeV}$~\cite{ParticleDataGroup:2026mpi} and all other quark masses and widths set to zero. This approximation induces only a small shift in the electroweak normalization and is negligible for the present hard coefficient study. The electromagnetic coupling is taken from the Fermi constant through
\begin{equation}
\alpha=\frac{\sqrt{2}}{\pi}G_\mu m_W^2\left(1-\frac{m_W^2}{m_Z^2}\right),
\end{equation}
which gives $\alpha=1/133.12$ for the inputs used here. The Cabibbo-Kobayashi-Maskawa matrix is set to the diagonal form. The NRQCD factorization scale $\mu_\Lambda$ is chosen following the standard practice in quarkonium phenomenology: it is taken of order the heavy-quark mass and varied between half and twice this value to estimate the residual factorization-scale dependence of the prediction~\cite{Bodwin:1994jh,Maltoni:2000km,Feng:2017hlu,Chen:2017pyi,Yu:2020tri,Sang:2020fql}. For the toponium system this prescription corresponds to the range $m_t/2 \leq \mu_\Lambda \leq 2\,m_t$, which stays well above the nonperturbative regime. The value of $\alpha_s(\mu_R)$ is obtained from the input $\alpha_s(m_Z)$ by solving the renormalization-group evolution in the same six-flavor scheme with \texttt{RunDec}~\cite{Herren:2017osy}.

The perturbative expansion is organized in terms of $\alpha_s(\mu_R)$ in the $\overline{\rm MS}$ scheme, with the top quark as the heavy degree of freedom matched onto the nonrelativistic theory. The renormalized NNLO short-distance coefficient defined in Eq.~\eqref{eq:kdef} is expanded as
\begin{equation}
K_{HZ}^{\mathrm{NNLO}}=1+\alpha_s(\mu_R)\;\delta K_1+\alpha_s^2(\mu_R)\;\delta K_2(\mu_R,\mu_\Lambda)\;,
\label{eq:nnlo-main}
\end{equation}
where $\delta K_{1,2}$ denote the NLO and NNLO correction factors of the partial decay width.
Our convention differs from that of Ref.~\cite{Kuhn:1995ee}, where the radiative correction is defined through a normalized decay-width ratio with the universal correction associated with the reference channel $\eta\to\gamma\gamma$ separated explicitly; the remaining finite contribution is denoted by $\delta k$. The two coefficients should not be identified directly; at NLO they are related by
\begin{equation}
\delta k=\frac{2\pi}{C_F}\,\delta K_1-\left(\frac{\pi^2}{2}-10\right).
\end{equation}
In the large-mass limit, defined as $m_X^2 \ll m_{\eta_t}^2$ with $X=H,Z$ and $M_{\eta_t}=2m_t$, the final-state boson masses are negligible and our full-mass result can be compared with the analytic result of Ref.~\cite{Kuhn:1995ee}, which was obtained in this limit. Taking the one-loop coefficient in this limit, $\delta K_1^{\rm LM}=-2.676$, and converting through the relation above, we reproduce $\delta k=-2-8\ln 2\simeq -7.55$ of Ref.~\cite{Kuhn:1995ee}. This provides a nontrivial check of the short-distance coefficient obtained in this work.

We now turn to the phenomenologically relevant case in which the full dependence on the Higgs- and $Z$-boson masses is retained. The NLO perturbative coefficient reads
\begin{equation}
\delta K_1 = -2.717\,,
\label{eq:coefficients:NLO}
\end{equation}
which is close to the large-mass-limit value $\delta K_1^{\rm LM}\simeq -2.68$, indicating that the mass dependence is mild at one loop. The exact mass dependence is nevertheless retained throughout the NNLO calculation.

The NLO coefficient contains no explicit $\mu_R$ or $\mu_\Lambda$ logarithm. At NNLO, the short-distance coefficient with finite mass takes the form
\begin{equation}
\begin{aligned}
\delta K_2(\mu_R,\mu_\Lambda)
=&-16.16\\
&-5.080\,\ln\frac{\mu_R^2}{m_t^2}
+3.567\,\ln\frac{\mu_R^2}{\mu_\Lambda^2}\,.
\label{eq:coefficients:NNLO}
\end{aligned}
\end{equation}
Compared with the NLO coefficient, Eq.~\eqref{eq:coefficients:NNLO} exhibits a nontrivial logarithmic structure involving both the renormalization scale $\mu_R$ and the NRQCD factorization scale $\mu_\Lambda$. The $\mu_R$-dependent logarithm originates from the renormalization of the short-distance amplitude and the running of the strong coupling, while the $\mu_\Lambda$ dependence arises from the matching onto the nonrelativistic operator and the factorization of the heavy-quarkonium matrix element. The latter dependence is not physical by itself and is compensated by the scale dependence of the corresponding long-distance matrix element, leaving the physical decay width independent of $\mu_\Lambda$ up to higher-order corrections.

The numerical result is also independent of the $\gamma_5$ prescription: carrying out the calculation in both the naive anticommuting scheme and the KKS prescription discussed in Sec.~\ref{subsec:amp}, we find identical short-distance coefficients after renormalization and matching. This confirms that the finite hard coefficient is insensitive to the treatment of $\gamma_5$ for the present non-anomalous contribution.

As a further check on the logarithmic structure of the NNLO coefficient, we verify that the explicit $\mu_R$ dependence in our result satisfies the renormalization-group constraint. Substituting
Eqs.~\eqref{eq:coefficients:NLO} and~\eqref{eq:coefficients:NNLO} into
Eq.~\eqref{eq:nnlo-main}, one obtains
\begin{equation}
\begin{aligned}
\mu_R^2\frac{d}{d\mu_R^2}K_{HZ}
=&\left(
\mu_R^2\frac{\partial}{\partial \mu_R^2}
-\frac{\beta_0}{4\pi}\alpha_s^2
\frac{\partial}{\partial \alpha_s}
\right)K_{HZ}\\
=&\mathcal O(\alpha_s^3)\;,
\end{aligned}
\end{equation}
where $\beta_0=11C_A/3-4T_F n_f/3$ with $T_F=1/2$ the normalization of the color generators. The cancellation occurs between the
explicit logarithms in the NNLO coefficient and the scale dependence generated by the running of the NLO term. Consequently, the residual
renormalization-scale dependence of the hard coefficient starts only at
$\mathcal O(\alpha_s^3)$, as expected for an NNLO result. This confirms that the scale logarithms in Eq.~\eqref{eq:coefficients:NNLO} are organized consistently with the renormalization-group structure of the short-distance coefficient.

\begin{figure}[tbp]
  \centering
  \includegraphics[width=0.48\textwidth]{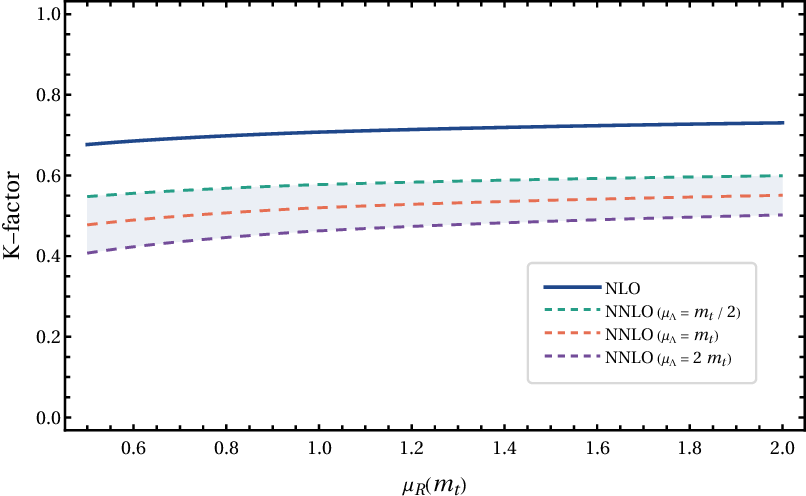}
  \caption{Renormalization-scale dependence of the full $K$ factor for $\eta_t\to HZ$. The solid curve shows the result truncated at NLO, while the dashed curves show the results truncated at NNLO for $\mu_\Lambda=m_t/2$, $m_t$, and $2\,m_t$. The shaded band represents the residual factorization-scale dependence of the NNLO prediction.}
  \label{fig:kfactor}
\end{figure}

Figure~\ref{fig:kfactor} presents the full $K$ factor defined in Eq.~\eqref{eq:nnlo-main} as a function of $\mu_R$ normalized to $m_t$; the curves distinguish the NLO prediction from the NNLO predictions obtained for three choices of the NRQCD factorization scale $\mu_\Lambda$. The variation with $\mu_R$ reflects the residual renormalization-scale dependence, whereas the spread among the $\mu_\Lambda$ choices indicates the residual factorization-scale dependence of the isolated hard coefficient.

For the default choice $\mu_R=\mu_\Lambda=m_t$, the logarithms in Eq.~\eqref{eq:coefficients:NNLO} vanish and the NNLO coefficient reduces to $\delta K_2=-16.16$. Although the absolute value of this coefficient is larger than that of the NLO coefficient in Eq.~\eqref{eq:coefficients:NLO}, its effect on the decay width is weighted by $\alpha_s^2(\mu_R)$. Taking $\alpha_s(m_t)\simeq 0.108$, the NLO and NNLO contributions to the $K$ factor are
\begin{align}
\alpha_s(m_t)\,\delta K_1 &\simeq -0.293\;,\\
\alpha_s^2(m_t)\,\delta K_2(m_t,m_t) &\simeq -0.188\;.
\end{align}
Therefore, the NLO correction reduces the leading-order width by about $29\%$, while the NNLO term gives an additional reduction of about $19\%$ at the central scale. The same sign of the NLO and NNLO corrections is noteworthy: the QCD radiative effects act coherently to suppress the leading-order width rather than cancel each other. The combined result is
\begin{equation}
K_{HZ}^{\rm NNLO}(m_t,m_t)\simeq 0.518\,.
\end{equation}
The relative size of the NNLO term, about two thirds of the NLO correction, indicates that the perturbative series of the hard kernel converges rather slowly; the NNLO term is therefore required for a reliable normalization of this channel in threshold studies.

The residual renormalization-scale uncertainty can be assessed by fixing the NRQCD factorization scale at its central value, $\mu_\Lambda=m_t$, and varying $\mu_R$ over $m_t/2$, $m_t$, and $2\,m_t$. Using the corresponding running coupling $\alpha_s(\mu_R)$, the NNLO terms contribute approximately $-19.5\%$, $-18.8\%$, and $-18.1\%$ to the leading-order width, respectively. After combining the NLO and NNLO contributions, the full NNLO $K$ factor changes from about $0.485$ to $0.548$ as $\mu_R$ is varied from $m_t/2$ to $2m_t$. This moderate variation reflects the residual short-distance perturbative uncertainty and results from the interplay between the explicit $\mu_R$ logarithms in the NNLO coefficient and the running of $\alpha_s(\mu_R)$. For $\mu_\Lambda=m_t/2$ this interplay cancels almost exactly: the NNLO contribution to $\delta K\equiv K_{HZ}-1$ varies only between $-0.127$ and $-0.132$ over the whole $\mu_R$ range, an apparent stabilization that should not be read as a preference for this particular factorization scale.

The dependence on the NRQCD factorization scale can be isolated by fixing $\mu_R=m_t$ and varying $\mu_\Lambda$ over the representative values $m_t/2$, $m_t$, and $2m_t$. Using $\alpha_s(m_t)\simeq0.108$, the NNLO contributions relative to the leading order width are approximately $-13.1\%$, $-18.8\%$, and $-24.6\%$ for $\mu_\Lambda=m_t/2$, $m_t$, and $2m_t$, respectively. Equivalently, the full NNLO $K$ factor changes from about $0.576$ to $0.460$ as $\mu_\Lambda$ is varied from $m_t/2$ to $2m_t$. The relatively large $\mu_\Lambda$ dependence should be interpreted as the residual factorization-scale dependence of the isolated hard coefficient, rather than as the uncertainty of a complete threshold prediction. In a full threshold-factorized calculation, this dependence is compensated by the scale dependence of the corresponding nonrelativistic matrix element or Green function.

\section{Summary}
\label{sec:summary}

We have computed the NNLO QCD correction to the hard short-distance coefficient for the pseudoscalar color-singlet threshold channel $\eta_t\to HZ$, retaining the full dependence on the Higgs- and $Z$-boson masses. The calculation includes the two-loop virtual corrections and the real double-gluon-emission channel. The result passes several internal checks: the one-loop coefficient reproduces the large-mass limit of Ref.~\cite{Kuhn:1995ee}, the extracted coefficient is identical in the two $\gamma_5$ prescriptions, and the explicit scale logarithms are organized consistently with the renormalization-group structure. At NLO the finite-mass coefficient stays close to its large-mass-limit value, while at NNLO it develops logarithms of both the renormalization scale $\mu_R$ and the NRQCD factorization scale $\mu_\Lambda$.

At the reference scale $\mu_R=\mu_\Lambda=m_t$, the NLO correction suppresses the leading-order width by about $29\%$, while the NNLO term gives a further suppression of about $19\%$, leaving $K_{HZ}^{\rm NNLO}\simeq0.52$. The two-loop contribution is about two thirds of the one-loop effect and has the same sign: the perturbative series of the hard kernel converges slowly, and the QCD radiative corrections act coherently to suppress the width. The residual $\mu_R$ dependence is moderate, while the $\mu_\Lambda$ dependence of the isolated coefficient is sizable; the latter is compensated, in a complete threshold prediction, by the evolution of the nonrelativistic matrix element or Green function and should not be read as the uncertainty of a physical prediction. Evolving the coefficient down to GeV-scale factorization scales even flips the sign of the NNLO term.

The hard coefficient obtained here is a necessary ingredient for a complete threshold prediction, but it is not by itself a physical observable.
A phenomenological description of $gg\to HZ$ near the $t\bar t$ threshold will require combining this NNLO kernel with the nonrelativistic Green function, finite-top-width effects, and a consistent matching to the continuum top-loop amplitude.
Such an analysis will show whether the toponium-like dynamics suggested by the recent LHC observations leaves a measurable trace in the $HZ$ final state, and how sharply this channel probes the top-Higgs interaction.

\begin{acknowledgments}
C.-P. Jia and Y.-Q. Ma is supported by the National Natural Science Foundation of China (No.~12325503) and the High-performance Computing Platform of Peking University. C.-P. Jia is also supported by the Young Faculty Research Capacity
Enhancement Program of Northwest Normal University (No.~20260004).
\end{acknowledgments}

\appendix

\section{Comparison of factorization-scale choices}

The numerical results in the main text are obtained with the NRQCD factorization scale chosen around the hard matching scale, $\mu_\Lambda\sim m_t$. For reference, we also compare them with the results obtained by taking $\mu_\Lambda$ at the GeV scale, between $1$ and $3~{\rm GeV}$. Such low values are sometimes used in conventional quarkonium applications, but for the top-quark system they lie far below the hard matching scale.

The comparison mainly serves as a diagnostic of the logarithmic $\mu_\Lambda$ dependence of the NNLO short-distance coefficient. When $\mu_\Lambda$ is lowered to the GeV scale, the NNLO correction changes substantially, reflecting the large logarithm generated by evolving the hard coefficient far away from its matching scale. In a complete threshold-factorized description, this dependence would be compensated by the evolution of the nonrelativistic matrix element or the Green function. The low-$\mu_\Lambda$ results should therefore be interpreted only as a scale-sensitivity check of the hard kernel, not as separate physical predictions.

\begin{figure*}[htbp]
  \centering
  \begin{subfigure}{0.48\textwidth}
    \centering
    \includegraphics[width=\textwidth]{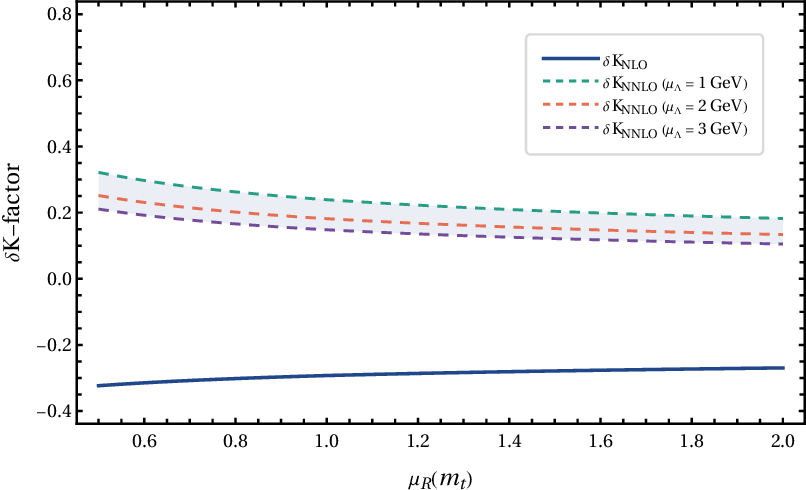}
    \caption{$\delta K$ for $\mu_\Lambda=1$, $2$, $3~{\rm GeV}$.}
  \end{subfigure}
  \hfill
  \begin{subfigure}{0.48\textwidth}
    \centering
    \includegraphics[width=\textwidth]{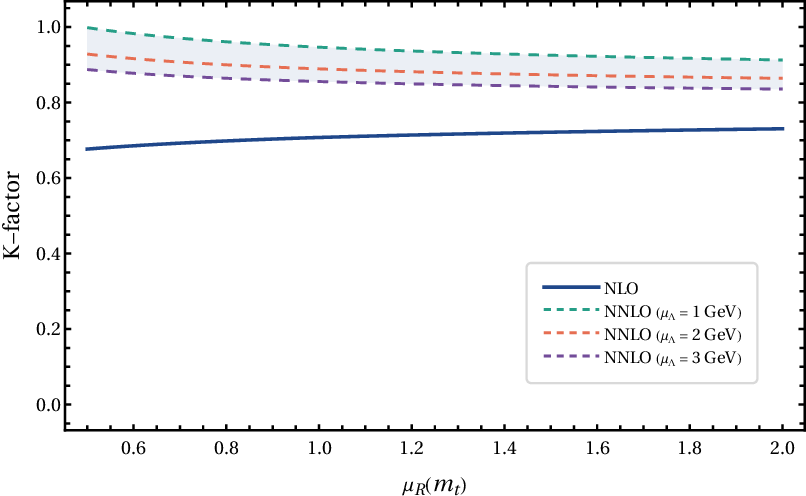}
    \caption{Full $K$ factor, same $\mu_\Lambda$ choices.}
  \end{subfigure}
  \caption{Comparison of factorization-scale choices for $\eta_t\to HZ$. The correction factor $\delta K=K_{HZ}-1$ and the full $K$ factor are shown as functions of $\mu_R/m_t$, with $\mu_\Lambda$ at the GeV scale.}
  \label{fig:deltakfactor-combined}
\end{figure*}

For this comparison we evaluate the NNLO correction factor of Eq.~\eqref{eq:nnlo-main} with the coefficients quoted to higher precision,
\begin{equation}
\begin{aligned}
\delta K_1 =& -2.71663\,,\\
\delta K_2(\mu_R,\mu_\Lambda)
=&-16.1579\\
&-5.0801\,\ln\frac{\mu_R^2}{m_t^2}
+3.56683\,\ln\frac{\mu_R^2}{\mu_\Lambda^2}\,.
\end{aligned}
\end{equation}
Figure~\ref{fig:deltakfactor-combined} shows the correction factor $\delta K\equiv K_{HZ}-1$ and the full $K$ factor as functions of $\mu_R/m_t$ for $\mu_\Lambda=1$, $2$, and $3~{\rm GeV}$. In this range the NNLO term flips sign relative to the $\mu_\Lambda\sim m_t$ results of Fig.~\ref{fig:kfactor} and largely cancels the NLO suppression, so the full $K$ factor stays close to unity, between about $0.83$ and $0.99$ over the $\mu_R$ interval considered. The spread among the three $\mu_\Lambda$ choices is moderate, of order $0.1$, as expected when the hard coefficient is evolved far below its matching scale and is dominated by the corresponding large logarithm.

\bibliographystyle{apsrev4-2}
\bibliography{etat_to_HZ_NNLO_QCD}

\end{document}